\documentclass[copyright,creativecommons]{eptcs}
\providecommand{\event}{FIT 2012} 

\pdfoutput=1

\title{A Parametric Counterexample Refinement Approach for Robust Timed Specifications}
\author{Louis-Marie Traonouez
\institute{Aalborg University, Denmark}
\email{lmtr@cs.aau.dk}
}
\def\titlerunning{Counterexample Refinement for Robust Timed Specifications}
\def\authorrunning{L.-M. Traonouez}

\usepackage[english]{babel}
\usepackage[utf8x]{inputenc}
\usepackage[T1]{fontenc}

\usepackage{amsmath,amssymb,stmaryrd,euscript,xspace,url,color,colortbl}

\usepackage{theorem}
\newtheorem{definition}{Definition}

\newtheorem{property}{Property}

\usepackage{etex,graphicx,caption,subfig,wrapfig,pgf,tikz}
\usetikzlibrary{arrows,automata,positioning}

\usepackage[english,lined,linesnumbered,ruled,nofillcomment]{algorithm2e}

\usepackage{multirow,pbox,ctable,booktabs}
\newcommand{\mc}[3]{\multicolumn{#1}{#2}{#3}}

\newcolumntype{K}{>{\columncolor[gray]{0.8}[.5\tabcolsep]}c}

\newcommand{\bbracket}[3] {\ensuremath{\llbracket{#1}\rrbracket^{#3}_{#2}} }
\newcommand{\ceil}[2][] {\ensuremath{\lceil{#2}\rceil_{#1}} }
\newcommand{\floor}[2][] {\ensuremath{\lfloor{#2}\rfloor_{#1}} }
\newcommand{\ie} {\textit{i.e.}\xspace}
\newcommand{\st} {\ensuremath{\ |\ }}
\newcommand{\sat}[1][] {\ensuremath{\ \mathbf{sat}_{#1}\ }}

\newcommand{\transition}[2] {\ensuremath{\smash{\xrightarrow{#1}}^{#2}}}
\newcommand{\univ} {\ensuremath{l_\mathrm{u}}}

\newcommand{\future} {\ensuremath{\mspace{-10.0mu}\nearrow}}
\newcommand{\past} {\ensuremath{\mspace{-5.0mu}\swarrow}}
\newcommand{\Post} {\textnormal{\textsf{Post}}\xspace}
\newcommand{\Pred} {\textnormal{\textsf{Pred}}\xspace}
\newcommand{\PPost} {\textnormal{\textsf{PPost}}\xspace}
\newcommand{\PPred} {\textnormal{\textsf{PPred}}\xspace}

\newcommand{\St} {\textit{St}}
\newcommand{\Loc} {\textit{Loc}}
\newcommand{\Clk} {\textit{Clk}}
\newcommand{\Act} {\textit{Act}}
\newcommand{\Acti} {\ensuremath{\textit{Act}_{\textnormal{\textrm{i}}}}\xspace}
\newcommand{\Acto} {\ensuremath{\textit{Act}_{\textnormal{\textrm{o}}}}\xspace}
\newcommand{\Outcome} {\ensuremath{\textnormal{\textsf{Outcome}}}}
\newcommand{\Bad} {\textnormal{\textsf{Bad}}\xspace}
\newcommand{\Runs} {\textnormal{\textsf{Runs}}\xspace}
\newcommand{\States} {\textnormal{\textsf{States}}\xspace}
\newcommand{\last} {\textnormal{\textsf{last}}\xspace}
\newcommand{\WSo} {\ensuremath{\textnormal{\textit{W}}^{\mathrm{o}}}\xspace}

\newcommand{\Inv} {\textit{Inv}}
\newcommand{\rob} {\textnormal{\textrm{rob}}}

\newcommand{\err} {\textnormal{\textsf{err}}}
\renewcommand{\phi} {\varphi}
\newcommand{\ssem}[1] {{\ensuremath{\llbracket{#1}\rrbracket_{\textrm{\textnormal{sem}}}}}}

\def\setR{{\mathbb R}} 
\def\setRp{\setR_{\geq 0}} 
\def\setRsp{\setR_{> 0}} 
\def\setQ{{\mathbb Q}} 
\def\setQp{\setQ_{\geq 0}} 
\def\setQsp{\setQ_{> 0}} 
\def\setN{{\mathbb N}} 
\newcommand{\vect}[1]{\mathbf{#1}}

\begin{document}

\maketitle

\begin{abstract}Robustness analyzes the impact of small perturbations in the semantics of a model.
		This allows to model hardware imprecision and therefore it has been applied to
		determine implementability of timed automata. In a recent paper, we extend this problem
		to a specification theory for real-timed systems based on timed input/output automata,
		that are interpreted as two-player games. We propose a construction that allows to synthesize 
		an implementation of a specification that is robust under a given timed perturbation,
		and we study the impact of these perturbations when composing different specifications.

		To complete this work we present a technique that evaluates the greatest admissible perturbation.
		It consists in an iterative process that extracts a spoiling strategy when a game is lost,
		and through a parametric analysis refines the admissible values for the perturbation. 
		We demonstrate this approach with a prototype implementation.
\end{abstract}

\section{Introduction}

Component-based design is a software development paradigm well established in the software engineering industry.  In component-based design, larger systems are built from smaller modules that depend on each other in well delimited ways described by interfaces.  The use of explicit interfaces encourages creation of robust and reusable components. \emph{Specification theories} provide a language for specifying component interfaces together with operators for combining them, such as parallel composition, along with algorithms for verification based on refinement checking.

For real-time systems, timed automata \cite{DBLP:journals/tcs/AlurD94} are the classical specification language.  Designs specified as timed automata are traditionally validated using model-checking against correctness properties expressed in a suitable timed temporal logic \cite{DBLP:journals/iandc/HenzingerNSY94}. Mature modeling and model-checking tools exist, such as Uppaal \cite{DBLP:journals/spe/BehrmannDLPY11}, that implement this technique and have been applied to numerous industrial applications.  

In \cite{David2010}, the authors proposed a specification theory for real time systems,
based on an input/output extension of timed automata model to specify both models and properties.
It uses refinement checking instead of model-checking to support compositionality
of designs and proofs from ground up.
The set of state transitions of the timed systems is partitioned between inputs,
representing actions of the environment, and outputs that represent the behaviour of the component.  The theory is equipped with a game-based semantic. The two players, Input and Output, compete in order to achieve a winning objective---for instance safety or reachability.

\begin{wrapfigure}{r}{4.5cm}
\centering
\includegraphics[trim = 5mm 25mm 0mm 20mm, clip, width=4cm]{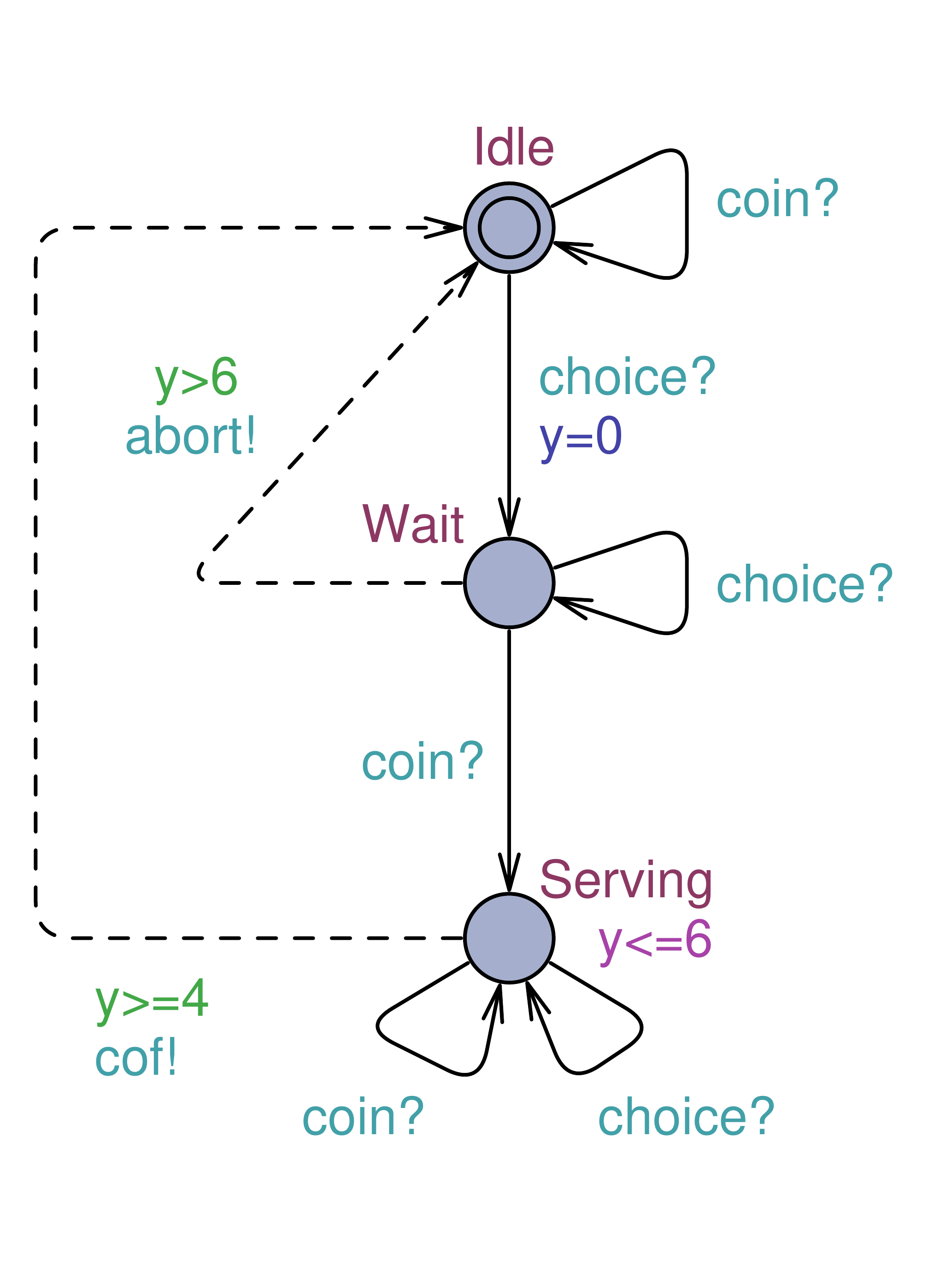}
\caption{Non robust specification of a coffee machine}
\label{fig_machine_non_rob}
\end{wrapfigure}

The theory of \cite{David2010} is equipped with a compatibility check and a consistency check that allows to decide whether a specification can indeed be implemented. Unfortunately, this check does not take limitations and imprecision of the physical world into account.  This is best explained with an example.  Consider the specification of a coffee machine in Fig.~\ref{fig_machine_non_rob}. This machine first ask for the choice of a drink, then awaits a coin, and after receiving the payment it delivers the coffee.  If the payment does not arrive 
within 6 time units, the machine aborts the drink selection and returns to the initial state, awaiting a new choice of a beverage. Already in this simple example it is quite hard to see,
that implementing a component satisfying this specification is not quite possible due to a subtle mistake. Observe that the two first steps of the machine are controlled by the environment, and not the system itself.  Thus any implementation has to be able to accept the following behaviour: first \textsf{choice?} and then the \textsf{coin?} arriving precisely 6 time units after the choice.  However then we arrive at the state $({\sf Serving},y=6)$ which requires that the coffee  (${\sf cof!}$) must be delivered immediately, in zero time.  No physical system would permit this, so we say that this state is not robustly consistent.

The above example can be fixed easily by adding another reset to clock \textsf{y}, when the \textsf{coin?} message is received.  It is probably the intended behaviour of the specification that the serving should take 6 time units from the insertion of the coin, and not from the choice of the drink.   Finding such errors in specifications is even harder in larger designs as non-robust timing can emerge in the compositions of multiple specifications, as a result of combing behaviours that themselves are robust.

The timing precision errors in specifications are not handled in any way in idealized interface theories such as \cite{David2010,DBLP:conf/emsoft/AlfaroHS02}. These and similar issues have let to a definition of the  so called timing \emph{robustness problem} that checks if a model can admit some timing perturbations while preserving a desired property. The robustness problem has been studied in various works for timed automata and it has been linked to the implementability problem  \cite{Wulf2005}.
In \cite{DBLP:conf/formats/LarsenLTW11}, we extend the specification theory of \cite{David2010} to support robustness analysis. We check robust consistency and robust compatibility under the assumption of a given small perturbation.
However, we were not able to decide if any perturbation can be admitted, neither determine the maximum amount.
That is the goal of this paper, to address the parametric problems for robust consistency and robust compatibility.
Our contributions include:
\begin{itemize}
\item We present a technique that evaluates the greatest admissible perturbation for the robustness problems.  We apply a counterexample abstraction refinement-like technique, that analyzes parametrically the results of lost timed games in order to refine the value of the perturbation.

\item We introduce a prototype tool that implements this technique and some other functionalities from the theory of \cite{DBLP:conf/formats/LarsenLTW11}.

\item We demonstrate the performances compared to a simple binary search technique for finding an optimal precision value.
\end{itemize}

\paragraph{Related works}

The robust semantics for timed automata with clock drifts has been introduced
by Puri\,\cite{Puri1998}.
The problem has been linked to the implementation
problem in \cite{Wulf2005}, which introduced the first semantics that modeled
the hardware on which the automaton is executed. In this work, the authors
proposed a robust semantics of Timed Automata called AASAP semantics (for
``Almost As Soon As Possible''), that enlarges the guards of an automaton by a
delay $\Delta$.
This work has been extended in \cite{Wulf2008} that proposes
another robust semantics with both clock drifts and guard enlargement.
Extending \cite{Puri1998} they solve the robust safety problem,
defined as the existence of a non-null value for the imprecision. They show
that in terms of robust safety the semantics with clock drifts is just as
expressive as the semantics with delay perturbation.

Robust timed games have been studied in \cite{DBLP:conf/formats/ChatterjeeHP08}.
In \cite{DBLP:conf/formats/LarsenLTW11}, we adapt their technique to check robust consistency and robust
compatibility.

Robustness is defined in \cite{Wulf2008} as the existence of a positive value for the imprecision of a timed automata. They prove that this problem is decidable, but they do not synthesize the value. A bound on the value is computed in \cite{BMS-formats11}.
Finally a quantitative analysis is performed in \cite{DBLP:conf/fossacs/JaubertR11} that computes the greatest admissible value for the perturbation, but the method is restricted to timed automata without nested loops. We propose an approximation technique that evaluates this value in the context of timed specifications, with no major restrictions on syntax of the specifications.

\paragraph{Organization of the paper: } We introduce in Section~\ref{sec:preliminaries} basic definitions for timed systems and timed games. In Section~\ref{sec:robust_specs} we recall the theory of robust timed specifications describe in  \cite{DBLP:conf/formats/LarsenLTW11} and \cite{David2010}.
The main contribution of this paper comes in Section~\ref{sec:cegar}, with a counterexample refinement technique to measure the imprecision allowed by the specifications. We present in Section~\ref{sec:impl} a tool that implements this technique,
and we demonstrate its performances in Section~\ref{sec:expe}.

\section{Preliminaries}\label{sec:preliminaries}

We use $\setN$ for the set of all non-negative integers,
$\setR$ for the set of all real numbers, and $\setRp$ (resp.\ $\setRsp$) for the non-negative
(resp.\ strictly positive) subset of $\setR$. Rational numbers are denoted by $\setQ$,
and their subsets are denoted analogously.


In the framework of \cite{David2010}, specifications and their implementations are semantically represented by Timed I/O Transition Systems (TIOTS) that are nothing more than timed transition systems with input and output modalities on transitions. 
Input represents the behaviours of the environment in which a specification is used, while output represents behaviours of the component itself. 

\begin{definition} A \emph{Timed I/O Transition System} is a tuple $S=
	(\St^S,s_0,\Sigma^S,\rightarrow^S)$, where $\St^S$ is an
	infinite set of states, $s_0 \in \St^S$ is the initial state,
	$\Sigma^S=\Sigma^S_i\oplus\Sigma^S_o$ is a finite set of actions
	partitioned into inputs $\Sigma^S_i$ and outputs $\Sigma^S_o$, and
	$\rightarrow^S : \St^S \times (\Sigma^S \cup \setRp) \times
	\St^S$ is a transition relation.
We write $s \transition{a}{S} s'$ when $(s,a,s')\in \rightarrow^S$ and
use $i?$, $o!$ and $d$ to range over inputs, outputs and $\setRp$, 
respectively.
\end{definition}

\noindent
In what follows, we assume that any TIOTS satisfies the following conditions:
\begin{itemize}
 \item time determinism: whenever $s \transition{d}{S} s'$ and $s \transition{d}{S} s''$ then $s' = s''$
 \item time reflexivity: $s \transition{0}{S} s$ for all $s \in \St^S$
 \item time additivity: for all $s,s'' \in \St^S$ and all $d_1,d_2 \in
	 \setRp$ we have $s \transition{d_1+d_2}{S} s''$ iff $s
	 \transition{d_1}{S} s'$ and $s' \transition{d_2}{S} s''$ for an $s'
	 \in \St^S$
\end{itemize}

\noindent
A {\em run} $\rho$ of a TIOTS $S$ from its state $s_1$ is a sequence $s_1 \transition{a_1}{S} s_2 \transition{a_2}{S} \dots \transition{a_{n}}{S} s_{n+1}$ such that for all $1 \leq i \leq n$,
$s_i \transition{a_{i}}{S} s_{i+1}$ with $a_i \in \Sigma^S \cup \setRp$.
We write ${\sf Runs}(s_1,S)$ for the set of runs of $S$ starting in $s_1$ and ${\sf Runs}(S)$ for ${\sf Runs}(s_0,S)$.
We write ${\sf States}(\rho)$ for the set of states reached in $\rho$, and if $\rho$ is finite ${\sf last}(\rho)$ is the
last state occurring in $\rho$.

A TIOTS $S$ is {\em deterministic} iff $\forall a\! \in \Sigma^S \cup \setRp$,
whenever $s \transition{a}{S} s'$ and $s \transition{a}{S} s''$, then $s'=s''$.
It is {\em input-enabled} iff each of its states $s \in \St^S$ is input-enabled:
$\forall i?\! \in \Sigma^S_i.\,\exists s'\!\in \St^S.\, s \transition{i?}{S} s'$.
It is {\em output urgent} iff $\forall s\!,s'\!,s''\! \in \St^S$ if $s \transition{o!}{S} s'$
and $s \transition{d}{S} s''$ then $d=0$.
Finally, $S$ verifies the {\em independent progress} condition iff
either $(\forall d\! \geq\! 0.\, s \transition{d}{S})$
or $(\exists d \!\in \setRp.\, \exists o!\! \in \Sigma^S_o.\, s \transition{d}{S} s'$ and $s' \transition{o!}{S})$.\\

TIOTS are syntactically represented by {\em Timed I/O Automata (TIOA)}.
Let $\Clk$ be a finite set of \emph{clocks}.
A \emph{clock valuation} over $\Clk$ is a mapping $\Clk \mapsto \setRp$ (thus $\setRp^\Clk$).
Given a valuation $u$ and $d \in \setRp$, we write $u+d$ for the valuation in which for each
clock $x \!\in \Clk$ we have $(u+d)(x) = u(x)+d$.
For $\lambda\subseteq\Clk$, we write $u[\lambda]$ for a valuation agreeing
with $u$ on clocks in $\Clk \setminus \lambda$, and mapping to $0$ the clocks in $\lambda$.

Let $\Phi(\Clk)$ denote all \emph{clock constraints} $\varphi$ generated by the grammar
$\varphi ::= x \prec k \st x-y \prec k \st \varphi \wedge \varphi$,
where $k\in \setQ$, $x,y \in \Clk$ and $\prec \,\in \{<,\leq,>,\geq\}$.
For $\varphi \in  \varPhi(\Clk)$ and $u \in \setRp^\Clk$, we write $u \models \varphi$ if $u$ satisfies $\varphi$.
Let $\bbracket{\varphi}{}{}$ denote the set of valuations $\{u \in \setRp^\Clk \st u \models \varphi\}$.
A subset $Z\subseteq\setRp^\Clk$ is a \emph{zone} if $Z=\bbracket{\varphi}{}{}$
for some $\varphi \in \varPhi(\Clk)$.

\begin{definition}\label{def:tioa}
  A \emph{Timed I/O Automaton} is a tuple ${\sf A} = (\Loc, q_0,\Clk,E,\Act,\Inv)$,
  where \Loc\ is a finite set of \emph{locations},
  $q_0 \in Loc$ is the \emph{initial location},
  $\Clk$ is a finite set of \emph{clocks},
  $E \subseteq \Loc \times \Act \times {\Phi}(\Clk) \times 2^{\Clk} \times \Loc$ is a set of \emph{edges},
  $\Act=\Acti\oplus \Acto$ is a finite set of \emph{actions},
  partitioned into \emph{inputs} ($\Acti$) and \emph{outputs} ($\Acto)$,
  $\Inv: \Loc \mapsto {\Phi}(\Clk)$ is a set of location \emph{invariants}.


We assume all TIOA include a \emph{universal location}, denoted $\univ$, that accepts every input and
can produce every output at any time.
\end{definition}

The semantics of a TIOA ${\sf A} = (\Loc,q_0,\Clk,E,\Act,\Inv)$ is a TIOTS $\ssem{\sf A}=(\Loc \times
\setRp^{\Clk},(q_0,\mathbf{0}),\linebreak[0] \Act,\rightarrow)$, where
$\mathbf{0}$ is a constant function mapping all clocks to zero, and
$\rightarrow$ is the largest transition relation generated by the
following rules:

\begin{itemize}
 \item Each edge $(q,a,\varphi,\lambda,q') \in E$ gives rise to $(q,u)
	 \transition{a}{} (q',u')$ for each clock valuation $u \in
	 \setRp^{\Clk}$ such that $u \models \varphi$ and $u'=
	 u[\lambda\mapsto 0]$ and $u' \models \Inv(q')$.
 \item Each location $q \in \Loc$ with a valuation $u \in \setRp^{\Clk}$ gives
	 rise to a transition $(q,u) \transition{d}{} (q,u+d)$ for each delay
	 $d \in \setRp$ such that $u +d \models \Inv(q)$.
\end{itemize}

\noindent Let $X$ be a set of states in $\ssem{\sf A}$ and let $a \in \Act$. The 
$a$-successors and $a$-predecessors of $X$ are defined respectively by:
$$\begin{array}{c}
  \Post_a(X)=\{(q',u') \st \exists (q,u) \in X.\, (q,u) \transition{a}{} (q',u')\}\\
  \Pred_a(X)=\{(q,u) \st \exists (q',u') \in X.\, (q,u) \transition{a}{} (q',u')\}
  \end{array}
$$
The timed successors and timed predecessors of $X$ are respectively defined by:
$$\begin{array}{c}
  X \future =\{(q,u+d) \st (q,u) \in X,\, d \in \setRp\}\\
  X \past =\{(q,u-d) \st (q,u) \in X,\, d \in \setRp\}
  \end{array}
$$
Additionally, we defined the safe timed predecessors of $X$ ${\it w.r.t}$ states $Y$,
that are the timed predecessors of $X$ that avoids the states of $Y$ along the path:
\begin{multline*}
\Pred_t(X,Y)=\{(q,u) \st \exists d \in \setRp.\, (q,u) \transition{d}{} (q,u+d) \text{ and } (q,u+d) \in X
\text{and } \forall d'\in [0,d].\, (q,u+d') \not\in Y\}
\end{multline*}

\paragraph{Symbolic Abstractions}

Since TIOTSs are infinite size they cannot be directly manipulated by computations. 
Usually {\em symbolic representations}, such as \emph{region graphs} \cite{DBLP:journals/tcs/AlurD94} or \emph{zone graphs},
are used as data structures that finitely represent semantics of TIOAs.
We denote by $X=(q,Z)$ a {\em symbolic state}, where $q \in \Loc$ and $Z \subseteq \setRp^\Clk$ is a zone.
The \emph{zone graph} is ${\sf G}_{\sf A}=({\EuScript Z}_{\sf A},X_0,\transition{}{})$,
where ${\EuScript Z}_{\sf A}$ is the set of reachable zones.
The initial state is defined by $X_0=\{(q_0,\vect{0})\} \future\cap\bbracket{\Inv(q_0)}{}{}$.
For $a \in \Act$, $(q,Z) \transition{a}{Z} (q',Z')$ if $(q,a,\varphi,\lambda,q') \in E$
and $Z'=((Z\cap\bbracket{\varphi}{}{})[\lambda])\future\cap\bbracket{\Inv(q')}{}{}$.

\begin{figure}[t]
\centering

\subfloat[Coffee machine]{
\label{fig_machine_spec}
\includegraphics[trim = 0mm 40mm 10mm 40mm, clip, width=3.5cm]{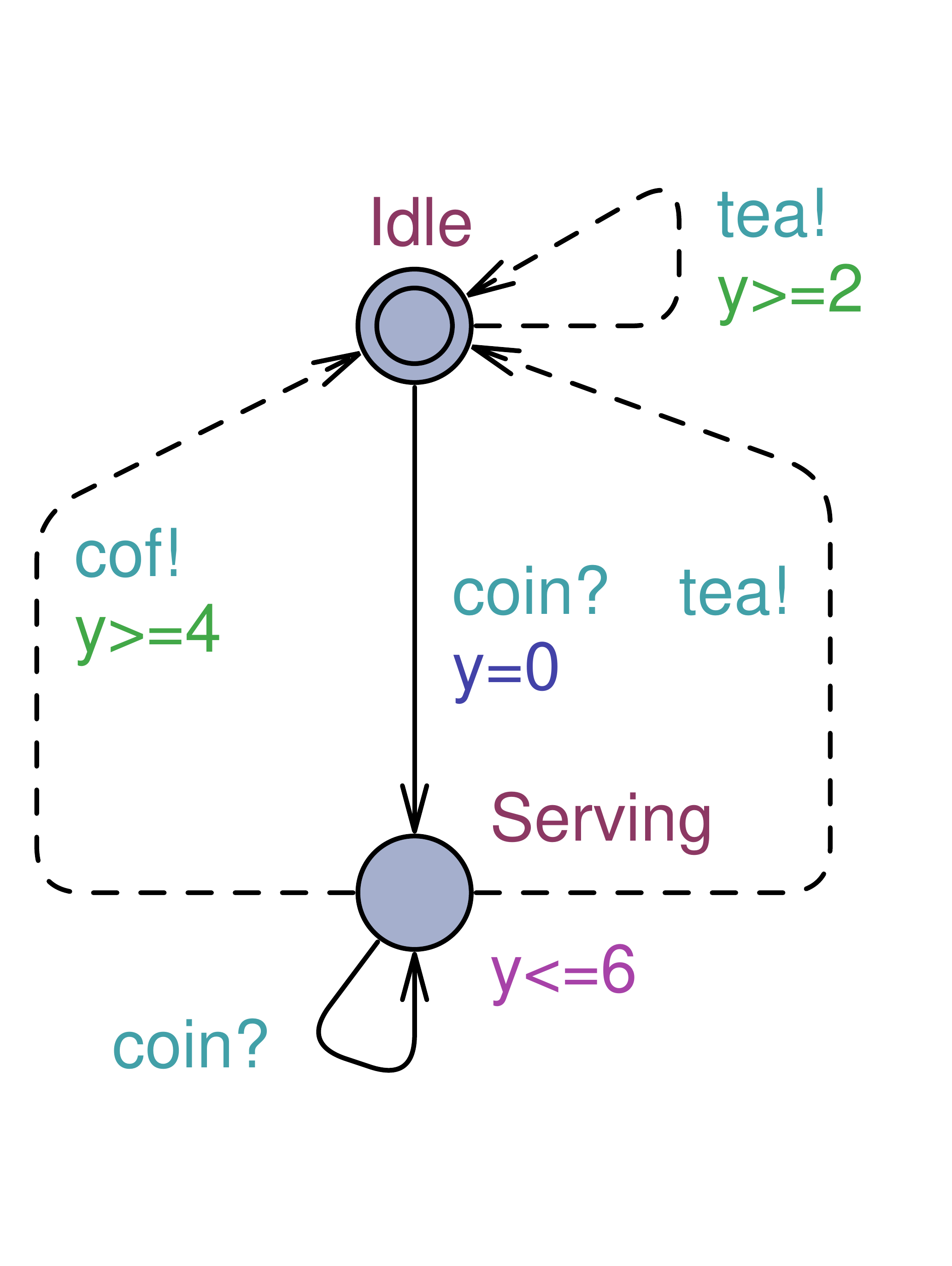}
}
\hspace{5mm}
\subfloat[Researcher]{
\label{fig_researcher_spec}
\includegraphics[trim = 0mm 75mm 10mm 70mm, clip, width=6.1cm]{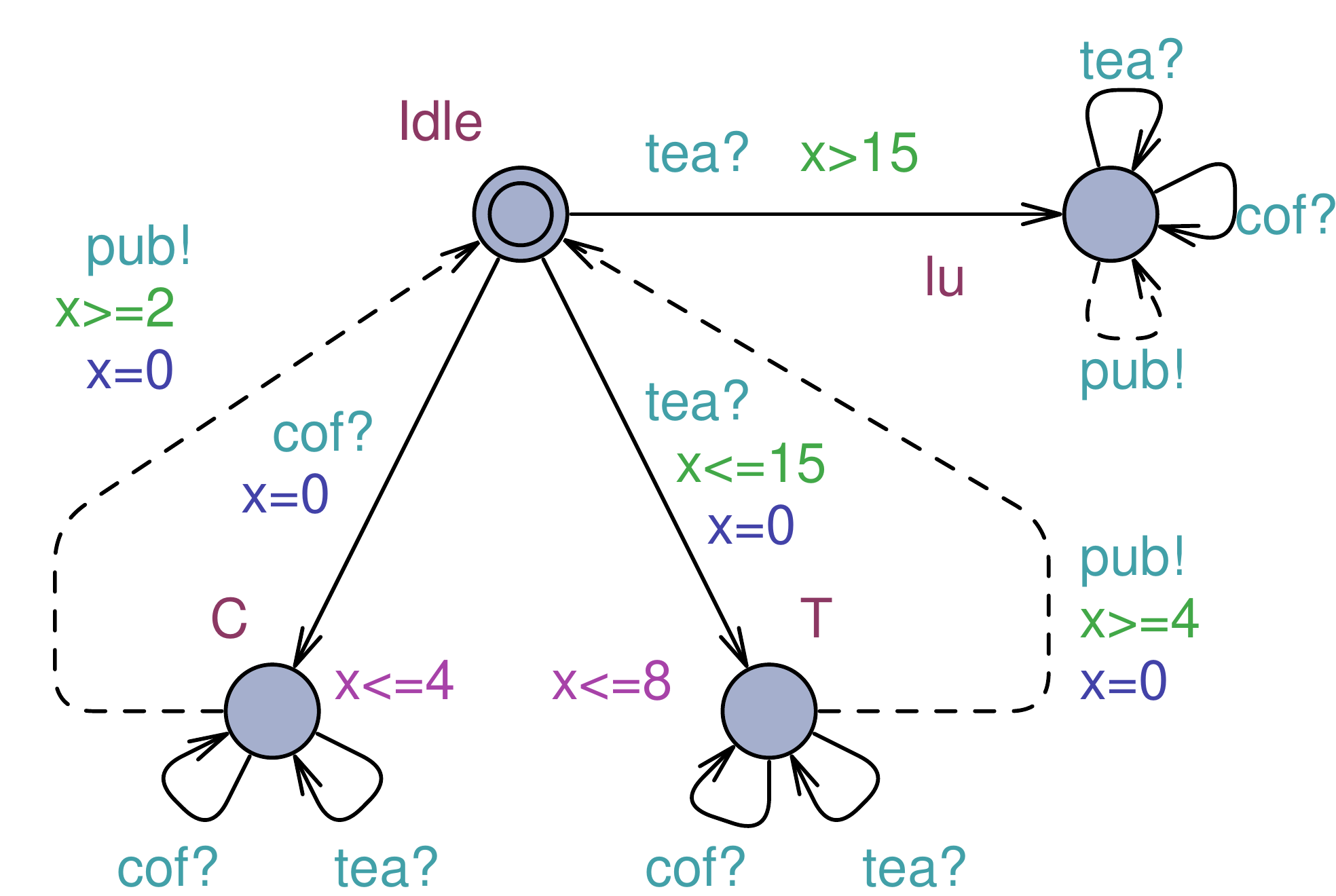}
}
\hspace{5mm}
\subfloat[Administration]{
\label{fig_administration_spec}
\includegraphics[trim = 10mm 45mm 10mm 45mm, clip, width=4.5cm]{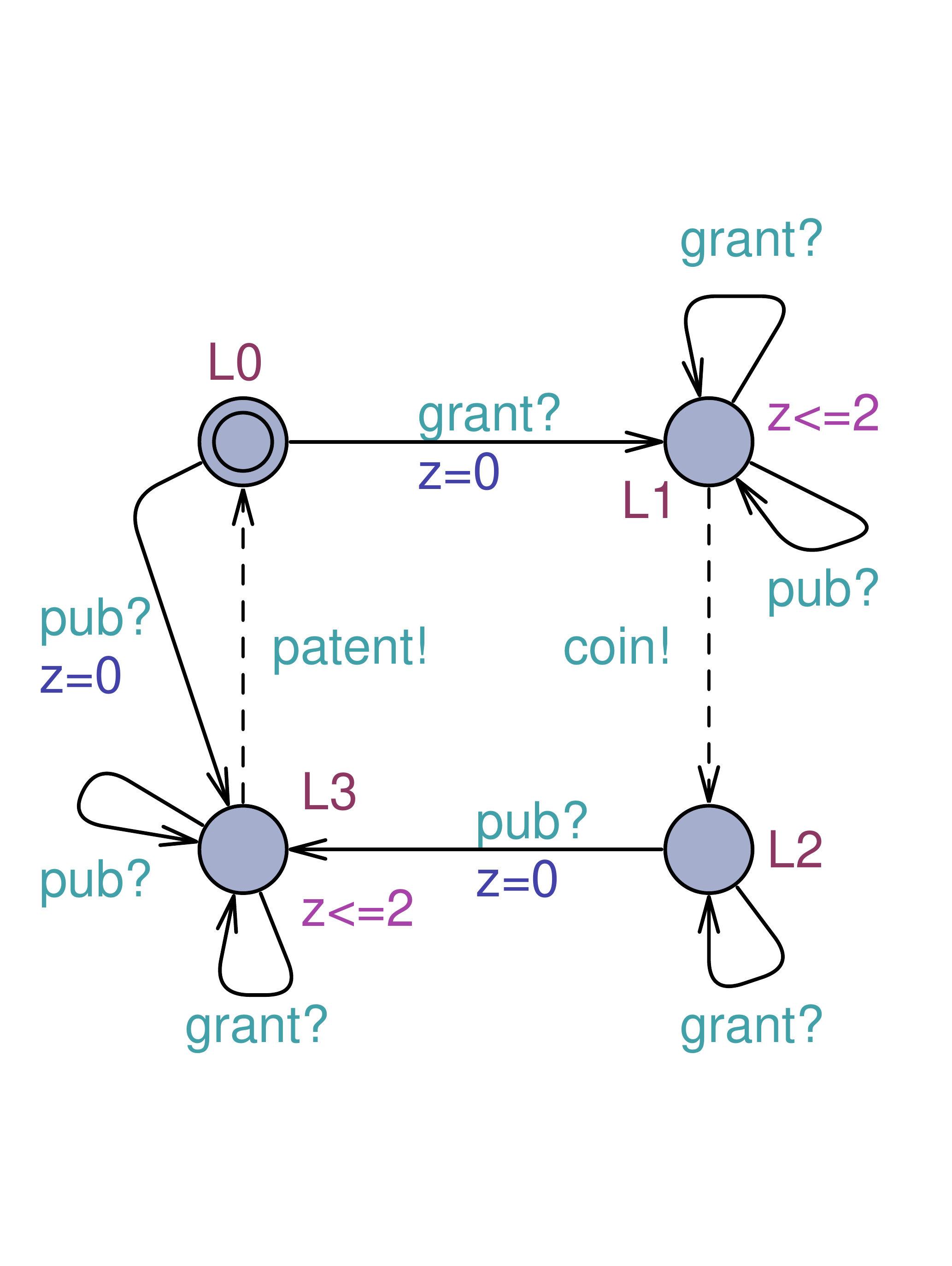}
}
\caption{Timed specifications with timed I/O automata}
\label{fig_tioa}
\end{figure}

\paragraph{Example} Figure~\ref{fig_tioa} presents three small examples of TIOAs, that specifies the behaviour of a university composed by
a coffee machine (Fig.~\ref{fig_machine_spec}), a researcher (Fig.~\ref{fig_researcher_spec})
and an administration (Fig.~\ref{fig_administration_spec}).

\paragraph{Timed Games}

TIOAs are interpreted as two-player real-time games between the \emph{output player} (the component)
and the \emph{input player} (the environment). The input plays with actions in \Acti and the output
plays with actions in \Acto. A strategy for a player is a function that defines her move at a certain time
(either delaying or playing a controllable action). 
A strategy is called \emph{memoryless} if the next move depends solely
on the current state. We only consider memoryless strategies, as these suffice for safety games
\cite{DBLP:conf/concur/AlfaroFHMS03}. For simplicity, we only define strategies for the output player
(i.e.\ output is the verifier). Definitions for the input player are obtained symmetrically.  
\begin{definition} 
\label{def:strategy}
	A memoryless \emph{strategy} $f_o$ for the output player on the TIOA $\sf A$
	is a partial function $\St^\ssem{\sf A} \mapsto \Acto \cup \{{\sf delay}\}$, such that
	\begin{itemize}
	 \item Whenever	$f_o(s) \in \Acto$ then $s \transition{f_o(s)}{} s'$ for some $s'$.
         \item Whenever $f_o(s) = {\sf delay}$ then $s \transition{d}{} s''$ for some $d>0$ and state $s''$, and $f_o(s'')={\sf delay}$.
	\end{itemize}
\end{definition}

\noindent The game proceeds as a concurrent game between the two player, each proposing its own strategy.
The restricted behaviour of the game defines the \emph{outcome} of the strategies.

\begin{definition} Let $\sf A$ be a TIOA, $f_o$ and $f_i$ be two strategies over $\sf A$ for the output and input player, respectively, and $s$ be a state of $\ssem{\sf A}$. $\Outcome(s,f_o,f_i)$ is the subset of ${\sf Runs}(s,\ssem{{\sf A}})$ defined inductively by:
\begin{itemize}
 \item $s \in \Outcome(s,f_o,f_i)$,
 \item if $\rho \in \Outcome(s,f_o,f_i)$, then $\rho' = \rho \transition{a}{} s' \in \Outcome(s,\linebreak[0]
      f_o,f_i)$ if $\rho' \in {\sf Runs}(s,\ssem{\sf A})$
       and one the following conditions hold:
  \begin{enumerate}
   \item $a \in \Act_o$ and $f_o({\sf last}(\rho)) = a$,
   \item $a \in \Act_i$ and $f_i({\sf last}(\rho)) = a$,
   \item $a \in \setRp$ and $\forall d \in [0,a[\, \exists s''.\, {\sf last}(\rho) \transition{d}{} s''$ and $\forall k \in \{o,i\}\,f_k(s'')={\sf delay}$.
  \end{enumerate}
 \item $\rho \in \Outcome(s,f_o,f_i)$ if $\rho$ infinite and all its finite prefixes are in $\Outcome(s,f_o,f_i)$.
\end{itemize} 
\end{definition}

\noindent
A \emph{winning condition} for a player in the TIOA $\sf A$ is a subset of $\Runs(\ssem{\sf A})$.
In safety games the winning condition is to avoid a set \Bad of ``bad'' states.
Formally, the winning condition is $\WSo(\Bad)=\{\rho \in \Runs(\ssem{\sf A}) \st \States(\rho)\linebreak[0] \cap \Bad = \emptyset \}$.
A strategy $f_o$ for output is a \emph{winning strategy} from state $s$ if and only if, for all strategy $f_i$ of input, $\Outcome_o(s,f_o,f_i) \subseteq \WSo(\Bad)$.
On the contrary, a strategy $f_i$ for input is a \emph{spoiling strategy} of $f_o$ if and only if $\Outcome(s,f_o,f_i) \not\subseteq \WSo(\Bad)$.
A state $s$ is winning for output if there exists a winning strategy from $s$. The game $({\sf A},\WSo(\Bad))$ is winning if and only if the initial state is winning.
Solving this game is decidable \cite{maler.ea:1995:stacs,CDFLL05,David2010}.
We only consider safety games in this paper, and without lost of generality we assume these ``bad'' states correspond to a set of entirely ``bad'' locations.

\paragraph{Symbolic Timed Games: } It is proved in \cite{DBLP:conf/concur/AlfaroFHMS03} that timed games can be solved 
using region strategies, where the players only need to remember the sequence of regions, instead of the sequence of states 
used in Definition~\ref{def:strategy}.
Consequently timed games can be solved through symbolic computations performed on the symbolic graph (either the region graph or the zone graph) using for instance the algorithm presented in \cite{CDFLL05}.
To represent these strategies we defined \emph{symbolic strategies} which apply on symbolic states:

\begin{definition}
	A \emph{symbolic strategy} $F_o$ for the output player on the symbolic graph 
	${\sf G}_{\sf A}=({\EuScript Z}_{\sf A},X_0,\transition{}{})$, 
	is a function ${\EuScript Z} \mapsto \Acto \cup \{{\sf delay}\}$,
	where  ${\EuScript Z}$ is a partition of the reachable states that refines ${\EuScript Z}_{\sf A}$,
	such that whenever $F_o((q,Z)) \in \Acto$ then $\forall u \in Z.\,(q,u) \transition{F_o((q,Z))}{} (q',u')$ for some $(q',u')$.
\end{definition}

\noindent We remark that a symbolic strategy $F_o$ corresponds to the set of strategies $f_o$ such that whenever $F_o((q,Z))=a$,
then $\exists u \in Z.\, f_o((q,u))=a$. 
For $(q,u) \in \ssem{\sf A}$, if $\exists Z.\, u \in Z \text{ and } F((q,Z)) \in \Act \cup \{{\sf delay}\}$,
we define by extension $F((q,u))=F((q,Z))$. For a symbolic state $X$ we define the timed successors of $X$ restricted by $F$ by:
\begin{multline*}
X\future^F=\{(q,u+d) \st (q,u) \in X,\, d\in \setRp,\, \forall d' \in [0,d].\\
F((q,u+d'))=F((q,u+d)) \vee F((q,u+d')) = \{\sf delay\}\}
\end{multline*}

\section{Robust Timed Specifications}\label{sec:robust_specs}

We summarize in this section the theory of robust timed specifications presented in \cite{DBLP:conf/formats/LarsenLTW11}.
It extends the theory of timed specifications based on TIOA presented in \cite{David2010}.

\subsection{Basics of the Timed Specification Theory}

In \cite{David2010} specifications and implementations are both represented by TIOAs satisfying additional conditions:
\begin{definition}
  A \emph{specification} ${\sf S}$ is a TIOA whose semantics
  $\ssem{\sf S}$ is deterministic and input-enabled.
\end{definition}
\begin{definition}  
  An \emph{implementation} ${\sf I}$ is a specification whose semantics
  $\ssem{\sf I\,}$ additionally verifies the output urgency and the
  independent progress conditions.
\end{definition}

In specification theories, a {\em refinement} relation plays a central role. It allows to compare specifications, and to relate implementations to specifications. In\,\cite{David2010}, as well as in\,\cite{DBLP:conf/sigsoft/AlfaroH01,alfaro/henzinger:2004,BulychevCDL09}, refinement is defined in the style of {\em alternating (timed) simulation}.
Formally, given two specifications $\sf S$ and $\sf T$, we say that
$\sf S$ \emph{refines} $\sf T$, written ${\sf S} \!\leq\! {\sf T}$, if and only if $\ssem{\sf S}$ is simulated by $\ssem{\sf T}$.
%
%
%
%
%

\begin{definition}\label{def:satisfaction}
  An implementation $\sf I$ \emph{satisfies} a specification $\sf S$, denoted $\,{\sf I} \sat {\sf S}$, if and only if
  $\,{\sf I\,}\leq {\sf S}$
\end{definition}

\noindent A specification  $\sf S$ is \emph{consistent} if and only if
there exists at least one implementation that satisfies $\sf S$.

A complete specification theory includes several operators to compose specifications.
The \emph{parallel composition} of two specifications $\sf S$ and $\sf T$
(denoted ${\sf S} \parallel {\sf T}$) is defined by the product of the two
TIOAs where components synchronize on common inputs/outputs.
Additional operators include conjunction and quotient. Their definition can be found in \cite{David2010}.

The parallel composition may introduce some incompatible states in the product,
\ie states in which the two components cannot work together.
With the input-enableness hypothesis no ``model-related'' errors can occurs when computing the product.
However specific incompatible states can be introduced in the models, by using for instance the universal location $\univ$
to specify an unpredictable behaviour of  the component.
A compatible environment for the two components allows to avoid these error states. 
We follow the optimistic approach of \cite{DBLP:conf/sigsoft/AlfaroH01},
\ie \emph{two specifications can be composed if there exists at least one environment 
in which they can work together}.
Formally, given a set ${\sf und}$ of undesirable states, we say that a specification $\sf S$ is \emph{useful} if there exists an 
environment $\sf E$ such that $\bbracket{{\sf S} \parallel {\sf E}\,}{sem}{} \cap {\sf und} = \emptyset$.
Two specifications $\sf S$ and $\sf T$ are {\em compatible} if and only if their product ${\sf S} \parallel {\sf T}$ is useful.

\subsection{Strategies in Timed Games as Operators on Timed Specifications}

The specification theory provides a game-based methodology in which winning strategies are used to synthesize implementations and compatible environments. Therefore, it determines consistency and usefulness of specifications,

In the \emph{consistency game} the output player tries to verify a safety condition,
\ie avoid a set of immediate inconsistent sates $\err^{\sf S}\subseteq \St^\ssem{\sf S}$.
Those are the sates that violate the independent progress condition:
\begin{equation*} 
	\err^{\sf S}\!=\!\left\{ s \,\big|\, (\exists d.\, s\not\!\!\transition{~d}{}) \text{ and } \forall d\, \forall o!\, \forall s'\!.\, s\transition{d}{}s' \text{ implies } s'\not\!\!\transition{~o!}{} \right\}
\end{equation*}
If output has a winning strategy $f_o$  in the timed game $({\sf S},\WSo(\err^{\sf S}))$, 
then one can synthesize from $f_o$ an implementation $\sf I$ of $\sf S$.

On the contrary in the \emph{usefulness game} the input player tries to avoid the set of incompatible states.
If there exists a winning strategy $f_i$ in the game $({\sf S},\WSo({\sf und}^{\sf S}))$,
it provides a compatible environment for $\sf S$.
This allows to prove usefulness of specifications and therefore compatibility between two specifications.

\subsection{Robust Implementations}

An essential requirement for an implementation is to be
realizable on a physical hardware, but this requires admitting small imprecisions characteristic for physical components
(computer hardware, sensors and actuators). 
The requirement of realizability has already been linked to the robustness
problem in \cite{Wulf2005} in the context of model checking.  
In specification theories the small deficiencies of hardware
can be reflected in a strengthened satisfaction relation, which introduces small perturbations to the timing of
implementation actions, before they are checked against the requirements of a specification---ensuring that the implementation
satisfies the specification even if its behaviour is perturbed.

We first formalize the concept of perturbation.
Let $\varphi \in \varPhi(\Clk)$ be a guard over the set of
clocks $\Clk$, let $x \in \Clk$ and $k \in \setQ$. The \emph{enlarged guard} $\ceil[\Delta]{\varphi}$
is constructed according to the following rules:
\begin{itemize}
\item Any term $x \prec k$ of $\varphi$ with $\prec\, \in\! \{<,\leq\}$ is replaced by $x \prec k \!+\! \Delta$

\item Any term $x \succ k$ of $\varphi$ with $\succ\, \in\! \{>,\geq\}$ is replaced by $x \succ k \!-\! \Delta$
\end{itemize}
Similarly, the \emph{restricted guard} $\floor[\Delta]{\varphi}$ is
using the two following rules:
\begin{itemize}
 \item Any term $x \prec k$ of $\varphi$ with $\prec \,\in\! \{<,\leq\}$ is replaced by $x \prec k \!-\! \Delta$

 \item Any term $x \succ k$ of $\varphi$ with $\succ \,\in\! \{>,\geq\}$ is replaced by $x \succ k \!+\! \Delta$.
\end{itemize}

\noindent Notice that for a for a clock valuation $u$ and a guard $\varphi$,
we have that $u \models \varphi$ implies $u \models \ceil[\Delta]{\varphi}$,
and  $u \models \floor[\Delta]{\varphi}$ implies $u \models \varphi$,
and $\floor[\Delta]{\ceil[\Delta]{\varphi}}=\ceil[\Delta]{\floor[\Delta]{\varphi}}=\varphi$.

We lift the perturbation to implementation TIOAs. Given a jitter $\Delta$,  the perturbation means a $\Delta$-enlargement of invariants and of output edge guards. Guards on the input edges are restricted by $\Delta$:
\begin{definition}\label{def:perturbation}
  For an implementation ${\sf I} \!=\! (\Loc,\linebreak[0] q_0,\linebreak[0] \Clk,\linebreak[0] E, \linebreak[0] \Act,\Inv)$ and $\Delta \!\in\! \setQsp$, the \emph{$\Delta$-perturbation} of ${\sf I}$ is the TIOA ${\sf I}_\Delta=(\Loc, q_0,\Clk,E',\Act,\Inv')$, such that:
\begin{itemize}
\item Every edge $(q,o!,\varphi,\lambda,q') \!\in\! E$ is replaced by $(q,o!,\linebreak[0]\ceil[\Delta]{\varphi},\linebreak[0] \lambda,q') \in E'$,
\item Every edge $(q,i?,\varphi,\lambda,q') \!\in\! E$ is replaced by $(q,i?,\linebreak[0]\floor[\Delta]{\varphi},\linebreak[0] \lambda,q') \in E'$,
\item $\forall q \in \Loc.\,\Inv'(q)=\ceil[\Delta]{\Inv(q)}$,
\item $\forall q \in \Loc.\,\forall i?\!\in\! \Acti$ there exists and edge $(q,i?,\varphi_\mathrm{u},\emptyset,\linebreak[0]\univ) \!\in\! E'$  with $\varphi_\mathrm{u}=\neg(\bigvee_{(q,i?,\varphi,\lambda,q')\in E} \floor[\Delta]{\varphi})$.
\end{itemize}
\end{definition}

\noindent ${\sf I}_\Delta$ is not necessarily action deterministic, as output guards are enlarged. However it is input-enabled, since by construction (last case in previous definition), any input not accepted after restricting input guards is redirected to the universal location $\univ$.  Also ${\sf I}_0$ equals ${\sf I}$.

In a similar manner, for a specification $\sf S$ we define $\ceil[\Delta]{\sf S}^\mathrm{o}$ the TIOA where all output edges and invariants have been enlarged.

%
%
%
\begin{definition}\label{def:robust-satisfaction}
  An implementation $\sf I$ robustly satisfies a specification $\sf S$ for a given delay $\Delta \in \setQp$, denoted ${\sf I} \sat[\Delta] {\sf S}$,
  if and only if\, ${\sf I}_\Delta \leq {\sf S}$
\end{definition}

\noindent A specification is {\em $\Delta$-robust consistent} if and only if it admits at least one $\Delta$-robust implementation.
A specification is {\em $\Delta$-robust useful} is there exists an environment $\sf E$, such that
$\ceil[\Delta]{\sf E}^\mathrm{o} \parallel {\sf S}$ avoids the errors states ${\sf und}^{\sf S}$.
As previously two specifications $\sf S$ and $\sf T$ are {\em $\Delta$-robust compatible} if and only if their composition
is $\Delta$-robust useful. The next property shows that robustness is monotonic for different values of the delay:

\begin{property}[Monotonicity]\label{prop:monotonic}
	Given two delays $0 < \Delta_1 \leq \Delta_2$ and an implementation $\sf I$: ${\sf I} \leq {\sf I}_{\Delta_1} \leq {\sf I}_{\Delta_2}$
	Therefore, if a specification $\sf S$ is $\Delta_2$-robust consistent, then $\sf S$ is also $\Delta_1$-robust consistent.
	Moreover if $\sf S$ is $\Delta_2$-robust useful, then $\sf S$ is $\Delta_1$-robust useful.
\end{property}

\subsection{Robust Timed Games for Timed Specifications}

Robust timed games add a robustness objective to safety games.
They can be used to verify robust consistency and robust compatibility, as it was done in the non-robust cases.
We have presented in \cite{DBLP:conf/formats/LarsenLTW11} a notion of robust strategies for timed games,
and we show how to synthesize robust implementations and robust environments from these strategies.
We finally give a construction of a robust game automaton,
whose original idea comes from \cite{DBLP:conf/formats/ChatterjeeHP08},
that transforms the original game.
It is shown that finding strategies in this automaton, using classical timed games algorithms, permits to synthesize
robust strategies in the original game. In this paper we always use with this construction to solve robust timed games.
Therefore we only recall its definition below:

\begin{definition}
\label{def:robust_automaton}
  Let $({\sf A},\WSo(\Bad))$ be a timed game, where ${\sf A}=(\Loc,q_0,\Clk,E,\Act,\linebreak[0]\Inv)$
  and $\Bad \in \Loc$, and let $\Delta \in \setQsp$.
  The  \emph{robust game automaton} ${\sf A}^\Delta_{\rob}=(\widetilde{\Loc},q_0,\Clk\cup
  \{y\},\widetilde{E},\linebreak[0]\Act\cup\{\rob\},\linebreak[0]
  \widetilde{\Inv})$ uses an additional clock $y$, and additional input action $\rob \in \Act_i$, and is constructed
  according to the following rules:
\begin{itemize}
\item $\Loc \subseteq \widetilde{\Loc}$, and for each location $q \in
  \Loc$ and each edge $e=(q,o!,\varphi,\lambda,q') \in E$, two
  locations $q^\alpha_{e}$ and $q^\beta_{e}$ are added in
  $\widetilde{\Loc}$. The invariant of $q$ is unchanged; the
  invariants of $q^\alpha_{e}$ and $q^\beta_{e}$ are $y \leq \Delta$.
\item Each edge $e'=(q,i?,\varphi,\lambda,q') \in E$ gives rise to the following edges in $\widetilde{E}$:\\
  $(q,i?,\varphi,\lambda,q')$,
  $(q^\alpha_e,i?,\varphi,\lambda,q')$ and
  $(q^\beta_e,i?,\varphi,\lambda,q')$.
\item Each edge $e=(q,o!,\varphi,\lambda,q') \in E$ gives rise to the following edges in $\widetilde{E}$:\\
   $(q,o!,\varphi,\{y\},q^\alpha_e)$,
   $(q^\alpha_e,o!,\{y=\Delta\},\{y\},q^\beta_e)$,
   $(q^\alpha_e,\rob,\varphi,\lambda,q')$,
   $(q^\beta_e,\rob,\varphi,\lambda,q')$,\\
   $(q^\alpha_e, \linebreak[0] \rob,\neg\varphi,\emptyset,\Bad)$ and $(q^\alpha_e,\rob,\neg\varphi,\emptyset,\Bad)$
   \footnote{Technically, since in a TIOA transitions guards must be convex, the last two transitions may be split into several copies, one for each convex guard in $\neg\varphi$.}
\end{itemize}
\end{definition}

\noindent The construction is demonstrated in Fig.~\ref{fig:transformation_robust_game}. 
The ideas behind the construction are that whenever output want to fire a transition $(q,o!,\varphi_o,\lambda_o,q_1)$ in the original automaton from a state $(q,u)$ after elapsing $d$ time units, this takes several steps in the robust automaton:
\begin{enumerate}
 \item Output proposes to play action $o!$ at time $d$ with the following sequence of transitions:
$$
(q,u)\transition{d-\Delta}{}(q,u+d-\Delta)\transition{o!}{}(q^\alpha,u+d-\Delta)\transition{\Delta}{}(q^\alpha,u+d)\transition{o!}{}(q^\beta,u+d)
$$
Note that this forbid output to play any action with a reaction time smaller than $\Delta$, and consequently this forbids Zeno strategies.
 \item Input can perturb this move with $d'\leq \Delta$, by choosing either a smaller delay:
  $$(q^\alpha,u+d-\Delta)\transition{d'}{}(q^\alpha,u+d-\Delta+d')\transition{\rob}{}(q_1,u+d-\Delta+d')$$
  or a greater delay: $$(q^\beta,u+d)\transition{d'}{}(q^\beta,u+d+d')\transition{\rob}{}(q_1,u+d+d')$$
 \item At any time in locations $q$,$q^\alpha$ and $q^\beta$, the original input edge $(q,i?,\varphi_i,\lambda_i,q_1)$ is still available.
 \item Output is implicitly forbidden to play a move that could not be perturbed since input will immediately win if the guard $\varphi_o$ is exceeded.
\end{enumerate}
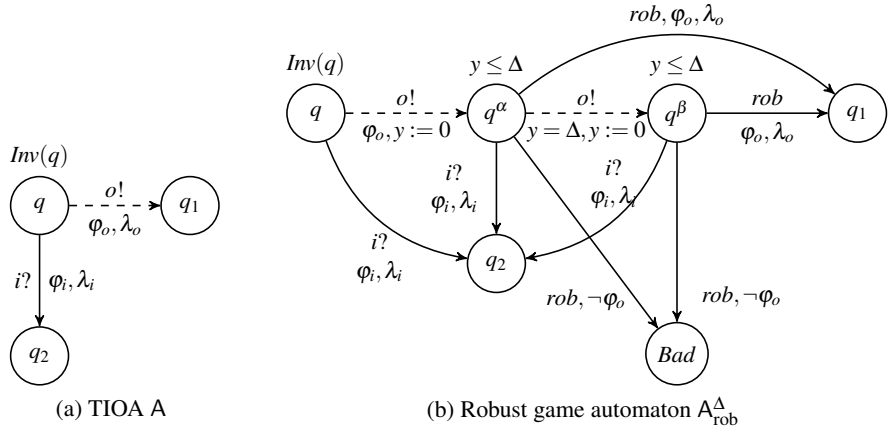
\begin{figure}[t]
\centering
\subfloat[TIOA $\sf A$]{
    \begin{tikzpicture}[transform shape,scale=0.80,->,>=stealth',node distance=2.5cm,auto,semithick]
      \tikzstyle{every state}=[draw=black,text=black]

      \node[state,label={above:$\Inv(q)$}] 		(q)	{$q$};
      \node[state,right of=q]				(q1)	{$q_1$};
      \node[state,below of=q]	(q2)	{$q_2$};

      \path (q)  edge[dashed]	node[label={above:$o!$},label={below:$\varphi_o,\lambda_o$}] {} (q1)
		 edge		node[label={left:$i?$},label={right:$\varphi_i,\lambda_i$}] {} (q2);
    \end{tikzpicture}
}
\hspace{5mm}
\subfloat[Robust game automaton ${\sf A}^\Delta_{\rob}$]{
    \begin{tikzpicture}[transform shape,scale=0.80,->,>=stealth',auto,node distance=3cm,semithick]
      \tikzstyle{every state}=[draw=black,text=black]

      \node[state,label={above:$\Inv(q)$}] (q)                       {$q$};
      \node[state,label={above:$y \leq \Delta$}] (qa) [right of=q] 	{$q^\alpha$};
      \node[state,label={above:$y \leq \Delta$}] (qb) [right of=qa]	{$q^\beta$};
      \node[state] (q1) [right of=qb] 	{$q_1$};
      \node[state,node distance=2.5cm] (q2) [below of=qa]       {$q_2$};
      \node[state,node distance=4cm] (qbad) [below of= qb] {$Bad$};

      \path (q) edge[dashed]           	node[label={above:$o!$},label={below:$\varphi_o,y:=0$}] {} (qa)
		edge[bend right]	node[label={below:$\begin{array}{c}
								i?\\
								\varphi_i,\lambda_i
							\end{array}$}] {} (q2);
      \path (qa) edge[dashed]          	node[label={above:$o!$},label={below:$y=\Delta,y:=0$}] {} (qb)
		 edge[bend left=40] 	node {$rob,\varphi_o,\lambda_o$} (q1)
		 edge[label distance=0.8cm]	node[label={-90:$rob,\neg\varphi_o$}] {} (qbad)
		 edge		     	node[label={left:$\begin{array}{c}
								i?\\
								\varphi_i,\lambda_i
							\end{array}$}] {} (q2);
      \path (qb) edge		        node[label={above:$rob$},label={below:$\varphi_o,\lambda_o$}] {} (q1)
		 edge[label distance=0.8cm]	node[label={-70:$rob,\neg\varphi_o$}] {} (qbad)
		 edge[bend left] 	node[label={90:$\begin{array}{c}
								i?\\
								\varphi_i,\lambda_i
							\end{array}$}] {} (q2);
    \end{tikzpicture}
}
\caption{Construction of the robust game automaton ${\sf A}^\Delta_{\rob}$ from an original automaton $\sf A$.}
\label{fig:transformation_robust_game}
\end{figure}

In \cite{DBLP:conf/formats/LarsenLTW11}, we prove that this construction is a sound technique to solve robust timed games and
check robust consistency and robust compatibility.

\section{Counter Strategy Refinement For Parametric Robustness}
\label{sec:cegar}

In previous section we have recalled our notions of robustness for a fixed delay.
In \cite{DBLP:conf/formats/LarsenLTW11} we additionally study the properties of these perturbations
with respect to the different operators in the specification theory.
In this paper we now consider the parametric problems, \ie determining the existence of a non-null delay. 
More precisely due to the monotonicity properties we would like to evaluate the greatest possible value of the perturbation.
The robustness problems that we consider in this section are the parametric extension of previously defined problems:
\begin{itemize}
 \item {\em Robust Consistency:} Given a specification $\sf S$, determine the greatest value of $\Delta$ such that $\sf S$ is $\Delta$-robust consistent.
 \item {\em Robust Usefulness:}  Given a specification $\sf S$, determine the greatest value of $\Delta$ such that $\sf S$ is $\Delta$-robust useful.
\end{itemize}

\subsection{Parametric Timed Games}

When we consider $\Delta$ as a free parameter, the robust game automaton construction of Section~\ref{sec:robust_specs} defines
a Parametric Timed I/O Automata, in a similar manner as Parametric Timed Automata are defined in \cite{DBLP:conf/stoc/AlurHV93,DBLP:journals/jlp/HuneRSV02}.
We denote by  $\varPhi_\Delta(\Clk)$ the set of parametric guards with parameter $\Delta$ over a set of clocks $\Clk$.
Parametric guards in $\varPhi_\Delta(\Clk)$ are generated by the following grammar $\varphi ::= x \prec l \st x\!-\!y \prec l \st \varphi \!\wedge\! \varphi$,
where $x,y \!\in\! \Clk$, $\prec \,\in \{<,\leq,>,\geq\}$ and $l=a+b*\Delta$ is a linear expression such that $a,b \in \setQ$.

\begin{definition}
A \emph{Parametric TIOA} with parameter $\Delta$, is a TIOA ${\sf A}$ such that guards and invariants are replaced by parametric guards.
\end{definition}

\noindent For a given value $\delta \in \setQp$, we define the non-parametric game ${\sf A}_{\delta}$ obtained by replacing each occurrence of the parameter $\Delta$ in the parametric guards of $\sf A$ by the value $\delta$. 

A parametric symbolic state $X$ is a set of triple $(q,u,\delta)$, where $\delta$ is a value of the parameter $\Delta$
and $(q,u)$ is a state in $\ssem{{\sf A}_{\delta}}$. Operations on symbolic states can be extended to parametric symbolic states,
such that  $X\future^P$, $X\past^P$,$\PPost_a(X)$, $\PPred_a(X)$ and $\PPred_t(X,Y)$ stands for the extensions of previously defined
non-parametric operations. Formally:
\begin{align*}
 X\future^P = & \{(q,u+d,\delta) \st (q,u,\delta) \in X,\, d \in \setRp\}\\
  X\past^P =   & \{(q,u-d,\delta) \st (q,u,\delta) \in X,\, d \in \setRp\}\\
  \PPost_a(X)=& \{(q',u',\delta) \st \exists (q,u,\delta) \in X.\, (q,u) \transition{a}{{\sf A}_{\delta}} (q',u')\}\\
  \PPred_a(X)=& \{(q,u,\delta) \st \exists (q',u',\delta) \in X.\, (q,u) \transition{a}{{\sf A}_{\delta}} (q',u')\}\\
  \PPred_t(X,Y)=& \{(q,u,\delta) \st \exists d \in \setRp.\, (q,u) \transition{d}{{\sf A}_{\delta}} (q,u+d)\\
	      & \text{ and } (q,u+d) \in X \text{ and } \forall d'\in [0,d].\, (q,u+d',\delta) \not\in Y\}
\end{align*}

\subsection{Parametric Robustness Evaluation}

Solving the robustness problems for any value of $\Delta$ would in general require to solve a parametric timed game.
This problem is undecidable as it has been shown that parametric model-checking problem is undecidable \cite{DBLP:conf/stoc/AlurHV93}.
In this paper, we propose to compute an approximation of the maximum delay perturbation.
Due to the monotonicity of the robustness problems (Property~\ref{prop:monotonic}),
we can apply an iterative evaluation procedure that searches for the maximum value until it belongs within a given precision interval.
This basic procedure is describe in Algorithm~\ref{algo:ref} for the parametric game $({\sf A}_\rob^\Delta,\WSo(\Bad))$
for output (again  it applies symmetrically to input).

\begin{algorithm}
\dontprintsemicolon
\KwIn{\begin{tabular}[t]{l}
	$({\sf A}_\rob^\Delta,\WSo(\Bad))$: parametric robust timed game,\\
	$\Delta_{\it max}$: initial maximum value,\\
	$\epsilon$: precision
      \end{tabular}
}
\KwOut{$\Delta_{\it good}$: maximum admissible value of $\Delta$}

\Begin{
  $\Delta_{\it good} \leftarrow 0$\;
  $\Delta_{\it bad} \leftarrow \Delta_{\it max}$\;
  
  \While{$\Delta_{\it bad}-\Delta_{\it good} > \epsilon$}{
    $(\Delta_{\it good},\Delta_{\it bad}) \leftarrow {\tt RefineValues}({\sf A}_\rob^\Delta,\Delta_{\it good},\Delta_{\it bad})$\;
  }
  \Return{$\Delta_{\it good}$}
}
\caption{Evaluation of parametric robustness }
\label{algo:ref}
\end{algorithm}

The algorithm assumes that the game $({\sf A}_\rob^0,\WSo(\Bad))$ is won, whereas the game $({\sf A}_\rob^{\Delta_{\it max}},\WSo(\Bad))$ is lost.
It verifies two invariants:  $\Delta_{\it good}$ stores the maximum value known to be correct  for the robust game; $\Delta_{\it bad}$ stores the minimum value known to be incorrect with precision $\epsilon$.
At the heart of the algorithm the procedure {\tt RefineValues} plays the game for a chosen value,
and update the variables $\Delta_{\it good}$ and $\Delta_{\it bad}$  according to the result.
Termination is ensure if each iteration reduces the length of the interval by some fixed minimum amount.

Different algorithms can be used to implement {\tt RefineValues}. A basic method is binary search.
In that case {\tt RefineValues} chooses the middle point $\Delta_{\it mid}$ of the interval $[\Delta_{\it good},\Delta_{\it bad}]$,
and  plays the game $({\sf A}_\rob^{\Delta_{\it mid}} ,\WSo(\Bad))$.
According to the results, it updates either $\Delta_{good}\it $ or $\Delta_{\it bad}$.
This algorithm has several drawbacks. First, the number of games it needs to solve heavily depends on the precision parameter.
Second, depending on the initial maximum value a high proportion of the games played may be winning,
and in that case the complete symbolic graph of the model must be explored.

\subsection{Counter Strategy Refinement}

We propose an alternative method that follows the principle of {\em counterexample-guided abstraction refinement}
\cite{DBLP:conf/cav/ClarkeGJLV00}.
In our settings, counterexamples are spoiling strategies computed when the game is lost.
We analyse these strategies in order to refine the value of $\Delta$.
Using this technique only the last game is winning.
The different steps are:
\begin{enumerate}
 \item Play the game $({\sf A}_\rob^{\Delta_{\it bad}} ,\WSo(\Bad))$.
 \item If the game is won, return the values $(\Delta_{\it bad},\Delta_{\it bad})$.    
 \item Else extract a counter strategy $F_i$ for the input player.
 \item Replay $F_i$ on the parametric game using Algorithm~\ref{algo:param_game}; it returns a value $\Delta_{\it min}$.
 \item If $\Delta_{\it min}$ is only an infimum and $\Delta_{\it bad}-\Delta_{\it min} > \epsilon$, return the values $(\Delta_{\it good},\Delta_{\it min})$.
 \item Else return the values $(\Delta_{\it good},\Delta_{\it min}-\epsilon)$.
\end{enumerate}

The goal of Algorithm~\ref{algo:param_game} is to replay the spoiling strategy $F_i$ on the parametric game
and compute the maximum value of $\Delta$ such that this strategy becomes infeasible.
It takes as inputs the parametric game automaton ${\sf A}_\rob^\Delta$,
the symbolic graph $({\EuScript Z}_{\sf A}^{\Delta_{\it bad}},X_0,\transition{}{})$ computed for the game $({\sf A}_\rob^{\Delta_{\it bad}},\WSo(\Bad))$,
and the spoiling strategy $F_i$.
It returns the infimum of the values $\Delta_{\it bad}$ such that $F_i$ is a spoiling strategy in the game $({\sf A}_\rob^{\Delta_{\it bad}},\WSo(\Bad))$.

The algorithm is similar to the timed game algorithm proposed in \cite{CDFLL05} and implemented in the tool TIGA \cite{DBLP:conf/cav/BehrmannCDFLL07}. However only the backward analysis is applied on parametric symbolic states,
starting from the ''bad'' locations. Additionally the algorithm only explores the states that belongs to the outcome of $F_i$.
Since $F_i$ is a spoiling strategy in a safety game, its outcome contains a set of finite runs that eventually reach the 
''bad'' locations. This ensures that a backward exploration restricted to this set of finite runs will terminate.
Formally, we define the outcome of symbolic spoiling strategy $F_i$ for input. 
$\Outcome(F_i)$ is the subset of runs in the symbolic graph defined inductively by:
\begin{itemize}
 \item $(q_0,S_0\future^{F_i}) \in \Outcome(F_i)$,
 \item if $\rho \in \Outcome(F_i)$ and $\last(\rho)=(q,Z)$, then $\rho' = \rho \transition{}{} (q',Z') \in \Outcome(F_i)$
       if $\exists (q,a,\varphi,\lambda,q') \in E$ and one of the following condition holds:
 \begin{enumerate}
   \item either $a \in \Act_i$ and $\exists Z''.F_i(Z'')=a$ and $Z'=\Post_a(Z \cap Z'')\future^{F_i}$,
   \item or $a \in \Act_o$ and $\exists Z''.F_i(Z'')={\sf delay}$ and $Z'=\Post_a(Z \cap Z'')\future^{F_i}$,
  \end{enumerate}
\end{itemize}
\noindent The backward exploration ends when the set of winning states $PWin[X_0]$ contains the initial state.
Then, the projection $(PWin[X_0]\cap \vect{0})_{|\Delta}$ computes the set of all the valuations of $\Delta$ such
that the strategy $F_i$ is winning. The algorithm returns the infimum of these valuations.

\begin{algorithm}[tb]
\dontprintsemicolon
\KwIn{\begin{tabular}[t]{l}
	$({\sf A}_\rob^\Delta,\WSo(\Bad))$: parametric robust timed game,\\
	$({\EuScript Z}_{\sf A}^{\Delta_{\it new}},X_0,\transition{}{})$: symbolic graph computed
	for the game $({\sf A}_\rob^{\Delta_{\it new}},\WSo(\Bad))$\\
	$F_i$: spoiling strategy for input
	in the game $({\sf A}_\rob^{\Delta_{\it new}},\WSo(\Bad))$
      \end{tabular}
}
\KwOut{Infimum of $\Delta_{\it bad}$ values such that $F_i$ is a spoiling strategy in $({\sf A}_\rob^{\Delta_{\it bad}},\WSo(\Bad))$}

\Begin{
  \tcc{Initialisation}
  $Waiting \leftarrow \emptyset$\;
  \For{$X=(q,Z) \in {\sf Z}_{\sf A}$}{
    \eIf{$q \in \Bad$}{
      $PWin[X] \leftarrow \bbracket{\Inv(q)}{}{}$\;
      $Waiting \leftarrow Waiting \cup \{Y \st \exists \rho.\,\rho \transition{}{} Y \transition{}{} X \in \Outcome(F_i)\}$\;
    }
    {
      $PWin[X] \leftarrow \emptyset$\;
    }
  }
  \tcc{Backward exploration}
  \While{$(Waiting \neq \emptyset) \wedge \vect{0} \not\in PWin[X_0])$}{
    $X=(q,Z) \leftarrow {\sf pop}(Waiting)$\;
    $PBad^* \leftarrow \neg\bbracket{\Inv(q)}{}{} \cup (\bigcup_{X \transition{a \in \Act_i}{} Y} \PPred_a(Win[Y]))$\;
    $PGood^* \leftarrow \bigcup_{X \transition{a \in \Act_o}{} Y} \PPred_a(\bbracket{\Inv(Y)}{}{} \setminus PWin[Y])$\;
    $PWin[X] \leftarrow \PPred_t(PBad^*,PGood^* \setminus PBad^*)$\;
    \BlankLine
    $Waiting \leftarrow Waiting \cup \{Y \st \exists \rho.\,\rho \transition{}{} Y \transition{}{} X \in \Outcome(F_i)\}$\;
  }
  \Return{${\sf Minimize}((PWin[X_0]\cap \vect{0})_{|\Delta})$}
}
\caption{Counter strategy refinement}
\label{algo:param_game}
\end{algorithm}

\section{Implementation}
\label{sec:impl}

The specification theory described in \cite{David2010} is implemented in the tool ECDAR \cite{DBLP:conf/atva/DavidLLNW10}.
In order to experiment the methods proposed in this paper, we have built a prototype in Python that reimplements the main functionalities
of ECDAR and support the analysis of the robustness of timed specifications \cite{pyECDAR}. Inside this tool, the theory presented in Section~\ref{sec:robust_specs} is implemented as a set of model transformations:
\begin{enumerate}
 \item Computation of ${\sf I}_\Delta$, the $\Delta$-perturbation of an implementation $\sf I$ for some $\Delta \in \setQp$.
 \item Computation of the robust game automaton ${\sf A}_\rob^\Delta$.
 \item In order to add rational perturbations on the models ${\sf I}_\Delta$ and ${\sf A}_\rob^\Delta$ the tool scales all the constants in the TIOA.
 \item Finally we transform the TIOA of a specification into a specific \emph{consistency game automaton} (resp. \emph{usefulness game automaton}), such that all non $\Delta$-robust consistent (resp. non $\Delta$-robust useful) states are observed by a single location.
\end{enumerate}
By combining these transformations we can check in the tool the three problems:
$\Delta$-robust satisfaction, $\Delta$-consistency and $\Delta$-usefulness.
The algorithms used are respectively the alternating simulation algorithm presented in \cite{CDFLL05}
and the on-the-fly timed games algorithm presented in \cite{BulychevCDL09}.

To solve the parametric robustness problems we have implemented the heuristic presented in Section~\ref{sec:cegar}
that approximates the maximum solution through a counter strategy refinement.
We have also implemented a binary search heuristic in order to compare the performances of the two approaches.
In Algorithm \ref{algo:param_game}, operations on parametric symbolic states are handled with the Parma Polyhedra Library \cite{ppl}.
We shall remark that using polyhedra increases the complexity of computations compared to Difference Bound Matrices (DBMs),
but this is necessary due to the form of the parametric constraints that are beyond the scope of classical DBMs.
This not so much a problem in our approach as parametric analysis is limited to spoiling strategies whose size is kept as small as possible.
Nevertheless an interesting improvement can be to use Parametric DBMs as presented in \cite{DBLP:journals/jlp/HuneRSV02}.

\section{Experiments}
\label{sec:expe}

We evaluate the performances of the tool to solve the parametric robustness problems on two academic examples.
We compare in these experiments the Counter strategy Refinement (CR) approach with the Binary Search (BS) method.
We presents benchmarks results for different values of the initial parameters $\Delta_{\it max}$ and $\epsilon$.

\subsection{Specification of a university}
The toy examples featured in this paper are extracted from \cite{David2010}. 
They describe the overall specification of a university, composed by three specifications: 
the coffee machine (M) of Fif.~\ref{fig_machine_spec}, the researcher (R) of Fig.~\ref{fig_researcher_spec}, and the administration (A)
of Fif.~\ref{fig_administration_spec}.
We study the robust consistency and the robust compatibility of these specifications and their parallel composition.
The results are presented in Tables~\ref{tab_univ_cons} and \ref{tab_univ_comp}.
The column \emph{game size} displays the size of the robust game automaton used in the analysis
in terms of locations (\emph{loc.}) and transitions (\emph{trans.}).
The next columns display the time spent to compute the maximum perturbation with different initial conditions.
The analysis of these results first shows that the Counter strategy Refinement method
is almost independent from the two initial parameters $\Delta_{\it max}$ and $\epsilon$.
This is not the case for Binary Search: the precision $\epsilon$ influences the number of games that must be solved,
and the choice of $\Delta_{\it max}$ change the proportion of games that are winning.
Comparing the results of the two methods shows that for most of the cases, especially the more complex one, the Counter strategy Refinement approach is more efficient.

\begin{table*}[tb]
\centering
\scriptsize
\setlength{\extrarowheight}{2pt}
\begin{tabular}{@{\hspace{0.3em}}c c@{\hspace{0.5em}}c K@{\hspace{0.8em}}c K@{\hspace{0.8em}}c K@{\hspace{0.8em}}c K@{\hspace{0.8em}}c@{\hspace{0.3em}}}
\toprule[0.1em]
			& 		&		& \mc{2}{c}{$\Delta_{\it max}=8$} & \mc{2}{c}{$\Delta_{\it max}=6$} & \mc{2}{c}{$\Delta_{\it max}=8$} & \mc{2}{c}{$\Delta_{\it max}=6$} \\
			& \mc{2}{@{\hspace{0.5em}}c}{\bf Game size}		& \mc{2}{c}{$\epsilon=0.1$}	  & \mc{2}{c}{$\epsilon=0.1$}	    & \mc{2}{c}{$\epsilon=0.01$}      & \mc{2}{c}{$\epsilon=0.01$} \\
\textbf{Model}		& loc.		& trans.	& CR	& BS 	 		  & CR		& BS 		    & CR 	& BS 	      	      & CR 	& BS 	\\
\midrule
$\mathbf{M}$  		& 9		& 21		& 119ms & 314ms			  & 119ms 	& 262ms		    & 119ms 	& 438ms		      & 119ms	& 437ms \\
$\mathbf{R}$		& 11		& 27		& 188ms & 303ms			  & 188ms	& 299ms		    & 188ms 	& 419ms		      & 188ms 	& 523ms \\
$\mathbf{A}$		& 9		& 22		& 133ms & 316ms			  & 133ms 	& 287ms		    & 133ms	& 441ms		      & 133ms 	& 483ms	\\
$\mathbf{M \parallel A}$& 41		& 158		& 10.1s	& 10.1s			  & 10.1s 	& 9.6s		    & 10.4s 	& 17.5s		      & 10.4s 	& 17.6s \\
$\mathbf{R \parallel A}$& 48		& 201		& 14.1s	& 12.1s			  & 12.5s 	& 11s		    & 14.1s 	& 19.6s		      & 12.5s 	& 19.4s	\\
$\mathbf{M \parallel R}$& 44		& 152		& 10s 	& 15.5s			  & 9.81s 	& 15.8s		    & 10.3s 	& 22.9s 	      & 9.78s 	& 29.2s	\\
$\mathbf{M  \parallel R \parallel A}$	& 180		& 803		& 54.4s & 56.3s			  & 54.6s 	& 112s		    & 55s 	& 58.8s		      & 55.7s 	& 216s	\\
\bottomrule
\end{tabular}
\caption{Robust consistency of the university specifications}
\label{tab_univ_cons}
\end{table*}

\begin{table*}[tb]
\centering
\scriptsize
\setlength{\extrarowheight}{2pt}
\begin{tabular}{@{\hspace{0.3em}}c c@{\hspace{0.5em}}c K@{\hspace{0.8em}}c K@{\hspace{0.8em}}c K@{\hspace{0.8em}}c K@{\hspace{0.8em}}c@{\hspace{0.3em}}}
\toprule
			& 			&		& \mc{2}{c}{$\Delta_{\it max}=8$} & \mc{2}{c}{$\Delta_{\it max}=6$} & \mc{2}{c}{$\Delta_{\it max}=8$} & \mc{2}{c}{$\Delta_{\it max}=6$} \\
			& \mc{2}{@{\hspace{0.5em}}c}{\bf Game size}			& \mc{2}{c}{$\epsilon=0.1$}	  & \mc{2}{c}{$\epsilon=0.1$}	    & \mc{2}{c}{$\epsilon=0.01$}      & \mc{2}{c}{$\epsilon=0.01$} \\
\textbf{Model}		& loc.		 	& trans. 	& CR	& BS 	 		  & CR		& BS 		    & CR 	& BS 	      	      & CR 	& BS 	\\
\midrule
$\mathbf{M \parallel R}$	& 21 			& 90		& 2.64s & 4.34s			  & 1.72s 	& 4.02s		    & 2.64s	& 5.5s		      & 1.72s 	& 5.45s\\
$\mathbf{M  \parallel R \parallel A}$	& 75			& 399		& 48s	& 65s			  & 42.7s	& 74.2s		    & 48.2s	& 78.1s		      & 42.9s	& 120s\\
\bottomrule
\end{tabular}
\caption{Robust compatibility between the university specifications}
\label{tab_univ_comp}
\end{table*}

\subsection{Specification of a Milner Scheduler}
The second experiment studies a real-time version of Milner's scheduler previously introduced in \cite{DBLP:conf/atva/DavidLLNW10}.
The model consists in a ring of $N$ nodes. Each nodes receives a start signal from the previous node to perform some work
and in the mean time forward the token to the next node within a given time interval.
We check the robust consistency of this model for different values of $N$ and different initial parameters.
The results are displayed in Table~\ref{tab_milner_cons}.
Like in previous experiment the results show that the Counter strategy Refinement method is independent form the initial conditions and in general more efficient than Binary Search.

\begin{table*}[tb]
\centering
\scriptsize
\setlength{\extrarowheight}{2pt}
\begin{tabular}{@{\hspace{0.3em}}c c@{\hspace{0.5em}}c K@{\hspace{0.8em}}c K@{\hspace{0.8em}}c K@{\hspace{0.8em}}c K@{\hspace{0.8em}}c@{\hspace{0.3em}}}
\toprule
		&		&		& \mc{2}{c}{$\Delta_{\it max}=30$} & \mc{2}{c}{$\Delta_{\it max}=31$} & \mc{2}{c}{$\Delta_{\it max}=30$} & \mc{2}{c}{$\Delta_{\it max}=31$} \\
		& \mc{2}{@{\hspace{0.5em}}c}{\bf Game size}	& \mc{2}{c}{$\epsilon=0.5$}	  & \mc{2}{c}{$\epsilon=0.5$}	    & \mc{2}{c}{$\epsilon=0.1$}       & \mc{2}{c}{$\epsilon=0.1$} \\
\textbf{Model}	& loc.		& trans.	& CR	& BS 	 		  & CR		& BS 		    & CR 	& BS 	      	      & CR 	& BS 	\\
\midrule
\textbf{1 Node}   & 13 & 35			& 0.97s & 0.68s 		  & 1.09s 	& 0.72s 	    & 0.97s 	& 1.03s 	      & 1.09s   & 1.09s\\
\textbf{2 Nodes}  & 81 & 344			& 10.7s & 10.3s			  & 11.2s 	& 12.6s		    & 10.5s 	& 15.8s 	      & 11.1s   & 19.4s \\
\textbf{3 Nodes}  & 449 & 2640			& 1m58  & 2m25 			  & 2m06	& 2m26	 	    & 1m57 	& 3m39		      & 2m05 	& 3m45 \\
\textbf{4 Nodes}  & 2305 & 17152		& 17m38 & 24m12			  & 17m38 	& 27m46 	    & 17m41	& 37m57 	      & 17m37	& 41m50 \\
\bottomrule
\end{tabular}
\caption{Robust consistency of Milner's scheduler nodes}
\label{tab_milner_cons}
\end{table*}

\subsection{Interpretation}

The performances of the Binary Search method depends on the number of games that are solved and on the outcome of these games.
Games that are winning (or games that are losing but with a value of $\Delta$ close to the optimum value) are harder to solve, since in these cases the (almost) complete symbolic state space must be explored.
Reducing the precision parameter $\epsilon$ implies that more games must be solved close to the optimum value, and therefore it increases the time of analysis.
Moreover, changing, even slightly, the initial maximum value $\Delta_{\it max}$ may change the number of games, but most important the outcome of these games, and therefore the proportion of winning games.
For instance in the last experiment, the expected result is $7.5$. With an initial value of 30 the bisections performed by the Binary Search method arbitrarily imply that only 1 game is winning out of 9 (for $\epsilon = 0.1$). With 31 this proportion is 6 out of 9, which increases the complexity of the analysis.

With the Counter strategy Refinement approach proposed in this paper only losing games are played until one is winning. The choice of $\Delta_{\it max}$ modifies the number of games that are solved, but in general the first games for large values of $\Delta$ are easily solved. Consequently, the choice of $\Delta_{\it max}$ shows in the experiments almost no impact on the performances.
With the parametric approach the parameter $\epsilon$ is only used when the value $\Delta_{\it min}$ computed by the refinement process is the minimum of the bad values. In that case the next iteration plays the game with the value $\Delta_{\it min} - \epsilon$. The experiments shows this has no impact on the performances.

\section{Conclusion}

We have studied the parametric robustness problems for timed specifications.
This works is based on the theory of timed specifications of \cite{David2010}.
It extends the theory of robust specifications of \cite{DBLP:conf/formats/LarsenLTW11},
which was limited to fix values for the delays.
More precisely, we evaluate through approximation techniques the maximum imprecision allowed by specifications.
To this end, we propose a counterexample refinement approach that analyses spoiling strategies in timed games.

This technique has been implemented in a prototype tool and its performances have been evaluated
during two experiments. The results show that our counterexample refinement technique offers
in most cases better and more robust ({\it w.r.t} initial conditions) performances than the binary search technique.
 
In a future version of our tool, we would like to apply the counterexample refinement approach to the alternating simulation game,
in order to solve the parametric satisfaction problem for an existing implementation.
We will also try to improve the performances; in particular for analysing parametric symbolic states.
An interesting approach could be to replace polyhedra by parametric DBMs.

\bibliographystyle{eptcs}
\bibliography{references}

\end{document}